\documentclass[aps,prb,twocolumn,showpacs,amssymb,floatfix]{revtex4-1}
\usepackage{amsmath,graphics,epsfig}
\usepackage{mathrsfs}
\usepackage{bm}

\newcommand{\tx}{\text}
\newcommand{\ti}{\textit}

\newcommand{\la}{\langle}

\newcommand{\nn}{\nonumber}

\newcommand{\ra}{\rangle}

\newcommand{\gm}{\gamma}

\newcommand{\Og}{\Omega}
\newcommand{\sg}{\sigma}

\newcommand{\ie}{{i.e.,~}}

\begin{document}
\title{Tailoring spin-orbit torque in diluted magnetic semiconductors}
\author{Hang Li}
\author{Xuhui Wang}
\author{Fatih Do\v{g}an}
\author{Aurelien Manchon}
\email[]{aurelien.manchon@kaust.edu.sa}
\affiliation{King Abdullah
University of Science and Technology (KAUST), Physical Science and
Engineering Division, Thuwal 23955-6900, Saudi Arabia}
\date{\today}

\begin{abstract}
We study the spin orbit torque arising from an intrinsic
\ti{linear} Dresselhaus spin-orbit coupling in a single layer
III-V diluted magnetic semiconductor. We investigate the transport
properties and spin torque using the linear response theory and we
report here : (1) a strong correlation exists between the angular
dependence of the torque and the anisotropy of the Fermi surface;
(2) the spin orbit torque depends nonlinearly on the exchange
coupling. Our findings suggest the possibility to tailor the spin
orbit torque magnitude and angular dependence by structural
design.
\end{abstract}
\pacs{72.25.Dc,72.20.My,75.50.Pp}
\maketitle

The electrical manipulation of magnetization is central to
spintronic devices such as high density magnetic random access
memory,\cite{Katine-prl-2000} for which the spin transfer torque
provides an efficient magnetization switching
mechanism.\cite{slonczewski-jmmm-1996,berger-prb-1996} Beside the
conventional spin-transfer torque, the concept of spin-orbit
torque in both metallic systems and diluted magnetic
semiconductors (DMS) has been studied theoretically and
experimentally.\cite{manchon-prb, garate-prb-2009, Hals-epl-2010,
chernyshov-nph-2009, endo-apl-2010, fang-nanotech-2011} In the
presence of a charge current, the spin-orbit coupling produces an
effective magnetic field which generates a non-equilibrium spin
density that in turn exerts a torque on the
magnetization.\cite{manchon-prb, garate-prb-2009, Hals-epl-2010}
Several experiments on magnetization switching in strained
(Ga,Mn)As have provided strong indications that such a torque can
be induced by a Dresselhaus-type spin-orbit coupling, achieving
critical switching currents as low as $10^{6}$
A/cm$^2$.\cite{chernyshov-nph-2009, endo-apl-2010,
fang-nanotech-2011} However, up to date very few efforts are
devoted to the nature of the spin-orbit torque in such a complex
system and its magnitude and angular dependence remain
unaddressed.

In this Letter, we study the spin-orbit torque in a diluted
magnetic semiconductor submitted to a linear Dresselhaus
spin-orbit coupling. We highlight two effects that have not been
discussed before. First, a strong correlation exists between the
angular dependence of the torque and the anisotropy of the Fermi
surface. Second, the spin torque depends \ti{nonlinearly} on the
exchange coupling. To illustrate the flexibility offered by DMS in
tailoring the spin-orbit torque, we compare the torques obtained
in two stereotypical materials, (Ga,Mn)As and (In,Mn)As.\par

The system under investigation is a uniformly magnetized single
domain DMS film made of, for example, (Ga,Mn)As or (In,Mn)As. We
assume the system is well below its critical temperature. An
electric field is applied along the $\hat{x}$ direction. It is
worth pointing out that we consider here a large-enough system to
allow us disregard any effects arising due to boundaries and
confinement.

We use the \ti{six-band} Kohn-Luttinger Hamiltonian to describe
the band structure of the DMS,\cite{fang-nanotech-2011}
\begin{align}
H_{\tx{KL}} = \frac{\hbar^2}{2m}
&\left[(\gamma_1+\frac{5}{2}\gamma_2)k^2
-2\gamma_3{(\bm{k}\cdot\hat{\bm{J}})^2}\right.\nn\\
&\left. +2(\gamma_3-\gamma_2)\sum_{i}k_{i}^2\hat{J}_{i}^2\right].
\label{eq:Ha}
\end{align}
where the phenomenological Luttinger parameters $\gamma_{1,2,3}$
determine the band structure and the effective mass of
valence-band holes. $\gamma_{3}$ is the anisotropy
parameter, $\hat{\bm{J}}$ is the total angular momentum and $k$ is the
wave vector. The bulk inversion asymmetry allows us to augment the Kohn-Luttinger
Hamiltonian by a strain-induced spin-orbit coupling of the
Dresselhaus type.\cite{chernyshov-nph-2009,garate-prb-2009} We assume
the growth direction of (Ga,Mn)As is directed along the $z$-axis,
two easy axes are pointed at $x$ and $y$,
respectively.\cite{Welp-prl-2003} In this case, the components of
the strain tensor $\epsilon_{xx}$ and $\epsilon_{yy}$ are
identical. Consequently, we may have a linear Dresselhaus spin-orbit
coupling\cite{chernyshov-nph-2009}
\begin{align}
H_{\tx{DSOC}}=\beta(\hat{\sigma}_x k_x-\hat{\sigma}_y k_y),
\label{eq:dsoc}
\end{align}
given $\beta$ the coupling constant that is a function of the
axial strain. \cite{chernyshov-nph-2009,bernevig-prb-2005}
$\hat{\sg}_{x(y)}$ is the $6\times6$ spin matrix of holes and $k_{x(y)}$ is the wave vector.

In the DMS systems discussed here,
we incorporate a mean-field like exchange coupling to
enable the spin angular momentum transfer between the hole
spin ($\hat{\bm{s}}=\hbar\hat{\bm\sg}/2$) and the localized ($d$-electron) magnetic moment
$\hat{\bm{\Omega}}$ of ionized $\tx{Mn}^{2+}$ acceptors,
\cite{abolfath-prb-2001,jungwirth-apl-2002}
\begin{align}
H_{\tx{ex}}=2J_{\tx{pd}}N_{\tx{Mn}}S_{a}\hat{\bm{\Omega}}\cdot{\hat{\bm{s}}}/\hbar
\end{align}
where $J_{\tx{pd}}$ is the antiferromagnetic coupling constant.
\cite{abolfath-prb-2001,bree-prb-2008} Here ${S_{a}=5/2}$ is the
spin of the acceptors. The hole spin operator, in the present
six-band model, is a $6\times 6$ matrix.\cite{abolfath-prb-2001}
The concentration of the ordered local Mn$^{2+}$ moments
$N_{\tx{Mn}}=4x/a^3$ is given as a function of $x$ that defines
the doping concentration of Mn ion. $a$ is the lattice constant.
Therefore, the entire system is described by the total Hamiltonian
\begin{align}
H_{\tx{sys}}=H_{\tx{KL}}+H_{\tx{ex}}+H_{\tx{DSOC}}.
\label{eq:total-hamiltonian}
\end{align}
In order to calculate the spin torque, we determine the
nonequilibrium spin densities $\bm{S}$ (of holes) as a \ti{linear}
response to an external electric field, \cite{garate-prb-2009}
\begin{equation}
\bm{S}=e E_x\frac{1}{V}\sum_{n,\bm{k}}
\frac{1}{\hbar\Gamma_{n,\bm{k}}}
\la\hat{\bm{v}}\ra\la \hat{\bm{s}} \ra \delta(E_{n,\bm{k}}-E_F).
\label{eq:spin-density}
\end{equation}
where $\hat{\bm{v}}$ is the velocity operator. In Eq.(\ref{eq:spin-density}), the scattering rate of hole
carriers by Mn ions is obtained by Fermi's
golden rule,\cite{jungwirth-apl-2002}
\begin{align}
\Gamma_{n,\mathbf k}^{\tx{Mn}^{2+}}=&\frac{2{\pi}}{\hbar}N_{\tx{Mn}}
\sum_{n^{\prime}}\int_{}\frac{d \bm{k}^{\prime}}{(2{\pi})^3}
{\left\vert M_{n,n^{\prime}}^{\bm{k},\bm{k}^{\prime}}\right\vert}^2\nonumber\\
&\times\delta(E_{n,\bm{k}}-E_{n^{\prime},\bm{k}^{\prime}})
(1-\cos\phi_{\bm{k},\bm{k}^{\prime}}),
\end{align}
where $\phi_{\bm{k},\bm{k}^{\prime}}$ is the angle between
two wave vectors $\bm{k}$ and $\bm{k}^{\prime}$. The matrix
element $M_{n,n^{\prime}}^{\bm{k},\bm{k}^{\prime}}$
between two eigenstates $(\bm{k},n)$ and
$(\bm{k}^{\prime},n^{\prime})$ is
\begin{align}
M_{n,n^{\prime}}^{\bm{k}\bm{k}^{\prime}}=& J_\tx{pd}S_a\la\psi_{n\bm{k}}
|\hat{\bm{\Omega}}\cdot{\hat{\bm{s}}}|\psi_{n^{\prime}\bm{k}^{\prime}}\ra\nonumber\\
&-\frac{e^{2}}{\epsilon({\left\vert\bm{k}-\bm{k}^{\prime}\right\vert}^{2}+p^{2})}\la\psi_{n\bm{k}}
\vert\psi_{n^{\prime}\bm{k}^{\prime}}\ra.
\label{eq:matrix-element}
\end{align}
Here $\epsilon$ is the dielectric constant of the host
semiconductors and $p=\sqrt{e^{2}g/\epsilon}$ is the Thomas-Fermi
screening wave vector, where $g$ is the
density of states at Fermi level. Finally, we calculate the field like spin-orbit torque
using\cite{manchon-prb}
\begin{equation}
\bm{T} = J_{\tx{ex}}\bm{S} \times \hat{\bm{\Og}},
\label{eq:spin-torque}
\end{equation}
where $J_{\tx{ex}}\equiv J_{\tx{pd}}N_{\tx{Mn}}S_{a}$. Throughout
this Letter, the results are given in terms of the torque
efficiency ${\bm T}/eE$. The interband transitions, arising from
distortions in the distribution function induced by the applied
electric field, are neglected in our calculation. This implies
that the torque extracted from the present model is expected to
accommodate only a field-like component.  The above protocols
based on linear response formalism allow us to investigate the
spin-orbit torque for a wide range of DMS material parameters.

\begin{figure}[tbh]
\begin{center}
\includegraphics[width=7cm]{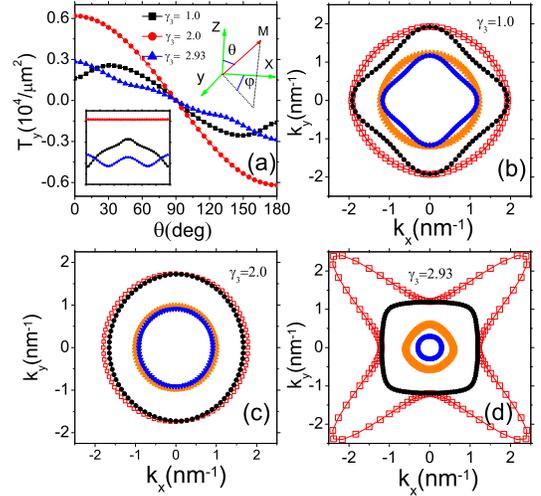}
\end{center}
\caption{(Color online) (a)The $y$-component of the spin torque as
a function of magnetization direction. Fermi surface intersection
in the $k_z=0$ plane for (b)$\gamma_3=1.0$, (c)$\gamma_3=2.0$ and
(c)$\gamma_3=2.93$. The red, black, orange and blue contours
stands for majority heavy hole, minority heavy hole, majority
light hole and minority light hole band, respectively. Inset (a)
depicts $T_y/\cos\theta$ as a function of magnetization direction.
The others parameters are $(\gamma_{1},\gamma_{2})=(6.98,2.0)$,
$J_{\tx{pd}}=55~\tx{meV}~ \tx{nm}^3$ and $p=0.2~\tx{nm}^{-3}$.}
\label{fig:st-gamma3}
\end{figure}

We plot in Fig.\ref{fig:st-gamma3}(a) the spin torque as a
function of the magnetization angle for different values of the
band structure  anisotropy parameter $\gamma_{3}$. The topology of
the Fermi surface can be modified by a linear combination of
$\gamma_{2}$ and $\gamma_{3}$: if $\gamma_{2}=\gamma_{3}\neq 0$,
the Fermi surface around the $\Gamma$ point is spherical, as shown
in Fig.\ref{fig:st-gamma3}(c). In this special case, the angular
dependence of the torque is simply proportional to $\cos\theta$
[red curve in Fig.\ref{fig:st-gamma3}(a)], as expected from the
symmetry of the $k$-linear Dresselhaus Hamiltonian, Eq.
(\ref{eq:dsoc})\cite{manchon-prb}. When $\gamma_3\neq\gamma_2$,
the Fermi surface deviates from a sphere
[Fig.\ref{fig:st-gamma3}(b) and (d)] and, correspondingly, the
angular dependence of the torque deviates from a simple
$\cos\theta$ function [\ie curves corresponding to
$\gamma_{3}=1.0$ and $\gamma_3=2.93$ in
Fig.\ref{fig:st-gamma3}(a)]. In a comparison to the spherical
case, the maximal value of the torque at $\theta=0$ is lower for
$\gm_{3}\neq\gm_{2}$. As Eq.(\ref{eq:spin-density}) indicates, in
the linear response treatment formulated here, the magnitude of
the spin torque is determined by the transport scattering time and
the expectation values of spin and velocity operators of holes.
Qualitatively, as the Fermi surface deviates from a sphere, the
expectation value $\la\hat{s}_{x}\ra$ of the heavy hole band,
contributing the most to the spin torque, is lowered at
$\theta=0$.

More specifically, as the Fermi surface warps, the angular
dependence of the spin torque develops, in addition to the
$\cos\theta$ envelop function, an oscillation with a period that
is shorter than $\pi$. The period of these additional oscillations
increases as the Fermi surface becomes more anisotropic in
$k$-space, see Fig. \ref{fig:st-gamma3}(b) and (d). To further
reveal the effect of band warping on spin torque, we plot
$T_{y}/\cos\theta$ as a function of the magnetization angle in
inset of Fig.\ref{fig:st-gamma3}(a). When $\gm_{3}=2.0$ (spherical
Fermi sphere), $T_{y}/\cos\theta$ is a constant, for $T\propto
\cos\theta$. When $\gm_{3}=2.93$ or $1.0$, the transport
scattering time of the hole carriers starts to develop an
oscillating behavior in $\theta$,\cite{rushforth-prl-2007} which
eventually contributes to additional angular dependencies in the
spin torque. The angular dependencies in spin-orbit torque shall
be detectable by techniques such as spin-FMR
\cite{fang-nanotech-2011}.

\begin{figure}[tbh]
\centering
\includegraphics[scale=.45]{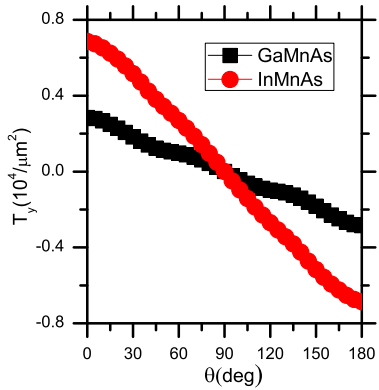}
\caption{(Color online) Torque $T_{y}$ as a function of the
magnetization direction for (Ga,Mn)As (black square) and (In,Mn)As
(red dots). For (Ga,Mn)As,
$(\gamma_1,\gamma_2,\gamma_3)=(6.98,2.0,2.93)$; for (In,Mn)As,
$(\gamma_1,\gamma_2,\gamma_3)=(20.0,8.5,9.2)$. The strength of the
spin-orbit coupling constant is: for (Ga,Mn)As, $\beta =
1.6~\tx{meV}~\tx{nm}$; for (In,Mn)As, $\beta =
3.3~\tx{meV}~\tx{nm}$.\cite{fabian-aps-2007} The exchange coupling
constant $J_{\tx{pd}}= 55~\tx{meV}~\tx{nm}^3$ for (Ga,Mn)As
\cite{ohno-jmmm-1999} and $39~\tx{meV}~\tx{nm}^3$ for
(In,Mn)As.\cite{wang-thesis-2002}} \label{fig:torque-both}
\end{figure}

In Fig.\ref{fig:torque-both}, we compare the angular dependence of
spin torque ($T_{y}$) for both (Ga,Mn)As and (In,Mn)As which are
popular materials in experiments and device fabrication.
\cite{Ohno-prl-1992,Koshihara-prl-1997,Jungwirth-prl-2002}
Although (In,Mn)As is, in terms of exchange coupling and general magnetic
properties, rather similar to (Ga,Mn)As, the difference
in band structures, lattice constants, and Fermi energies between
these two materials gives rise to different density of states,
strains, and transport scattering rates. For both materials, the
spin torque decrease monotonically as the angle $\theta$ increases
from $0$ to $\pi/2$. Throughout the entire angle range $[0,\pi]$,
the amplitude of the torque in (In,Mn)As is twice larger than that in
(Ga,Mn)As. We mainly attribute this to two effects. First of all,
the spin-orbit coupling constant $\beta$ in (In,Mn)As is about
twice as larger than that in (Ga,Mn)As. Second, for the same hole
concentration, the Fermi energy of (In,Mn)As is higher than that
of (Ga,Mn)As.

\begin{figure}[tbh]
\begin{center}
\includegraphics[width=6cm]{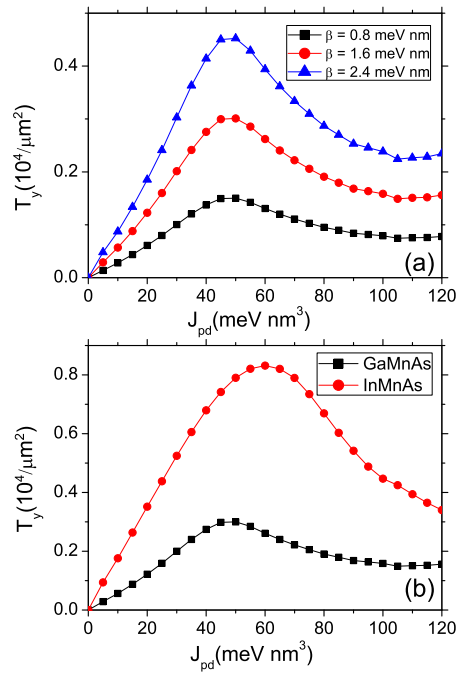}
\end{center}
\caption{(Color online) The $T_y$ component of the spin torque
as a function of exchange coupling $J_{\tx{pd}}$.
(a) $T_{y}$ versus $J_{\tx{pd}}$ at various
values of $\beta$, for (Ga,Mn)As. (b) $T_{y}$ versus
$J_{\tx{pd}}$, for both (Ga,Mn)As and (In,Mn)As.
The magnetization is directed along the $z$-axis ($\theta=0$).
The other parameters are the same as those in Fig.\ref{fig:torque-both}.}
\label{fig:st-exchange}
\end{figure}

In the following, we further demonstrate a
counter-intuitive feature that, in the DMS system considered in
this Letter, the spin orbit torque depends nonlinearly on the
exchange splitting. In Fig. \ref{fig:st-exchange}(a), $T_y$
component of the spin torque is plotted as a function of the
exchange coupling $J_{\tx{pd}}$, for different values of $\beta$.
In the weak exchange coupling regime, the electric generation of
non equilibrium spin density dominates, then the leading role of
exchange coupling is defined by its contribution to the transport
scattering rate. We provide a simple qualitative explanation on
such a peculiar $J_{\tx{pd}}$ dependence. Using a Born
approximation, the scattering rate due to the $p-d$ interaction is
proportional to $1/\tau_{J}=b J_{\tx{pd}}^{2}$, where parameter
$b$ is $J_{\tx{pd}}$- independent. When the nonmagnetic scattering
rate $1/\tau_{0}$ is taken into account, \ie the Coulomb
interaction part in Eq.(\ref{eq:matrix-element}), the total
scattering {time} in Eq.(\ref{eq:spin-density}) can be estimated
as
\begin{align}
\frac{1}{\hbar\Gamma}\propto
\frac{1}{bJ_{\tx{pd}}^{2}+\frac{1}{\tau_{0}}},
\label{eq:scattering-rate-estimate}
\end{align}
which contributes to the torque by $T\propto
J_{\tx{pd}}/(\hbar\Gamma)$. This explains the transition behavior,
\ie increases linearly then decreases, in the moderate
$J_{\tx{pd}}$ regime in Fig.\ref{fig:st-exchange}. As the exchange
coupling further increases, Eq.(\ref{eq:scattering-rate-estimate})
is dominated by the spin-dependent scattering, therefore the
scattering time $1/\hbar\Gamma \propto 1/J_{\tx{pd}}^{2}$.
Meanwhile, the energy splitting due to the exchange coupling
becomes significant, thus $\la\hat{\bm{s}}\ra\propto J_{\tx{pd}}$.
In total, the spin torque is insensitive to $J_{\tx{pd}}$,
explaining the flat curve in the large exchange coupling regime.
In Fig. \ref{fig:st-exchange}(b), we plot the influence of the
exchange coupling on the spin torque for two materials. In
(In,Mn)As, mainly due to a larger Fermi energy in a comparison to
(Ga,Mn)As, the peak of the spin torque shifts towards a larger
$J_{\tx{pd}}$. The dependence of the torque as a function of the
exchange in (In,Mn)As is more pronounced than in (Ga,Mn)As, due to
a stronger spin-orbit coupling.

\begin{figure}[tbh]
\begin{center}
\includegraphics[width=6cm]{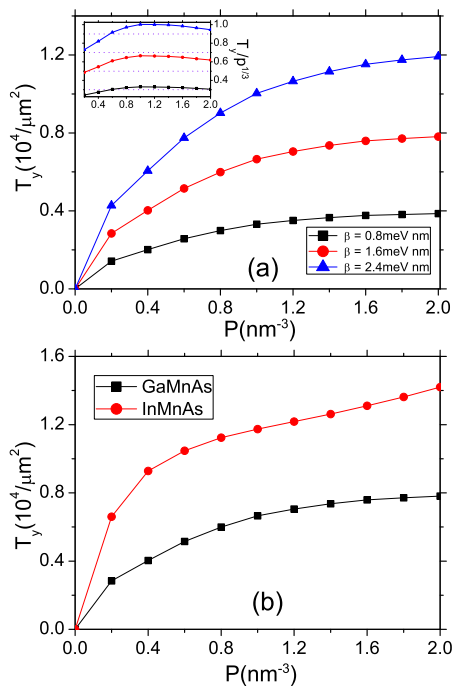}
\end{center}
\caption{(Color online) The y-component of the spin torque
as a function of hole concentration.
(a) The y-component of the spin torque versus hole concentration
at different $\beta$. (b) spin torque versus hole concentration in (Ga,Mn)As
and (In,Mn)As. For (Ga,Mn)As, $J_{\tx{pd}}=55~\tx{meV}~\tx{nm}^3$; for (In,Mn)As,
$J_{\tx{pd}}=39~\tx{meV}~\tx{nm}^3$.
The other parameters are the same as in Fig.\ref{fig:st-exchange}.}
\label{fig:st-hole}
\end{figure}

The possibility to engineer electronic properties by doping is
one of the defining features that make DMS promising for
applications. Here, we focus on the doping effect which allows the
spin torque to vary as a function of hole carrier concentration.
In Fig. \ref{fig:st-hole}(a), the torque is plotted as a function of
the hole concentration for different $\beta$ parameters. With the
increase of the hole concentration, the torque increases due to
an enhanced Fermi energy. In the weak spin-orbit
coupling regime (small $\beta$), the torque as a function of the hole concentration
($p$) follows roughly the $p^{1/3}$ curve as shown in the inset
in Fig. \ref{fig:st-hole}(a).
The spherical Fermi sphere approximation and a simple parabolic dispersion
relation allow for an analytical expression of the spin torque,
\ie in the leading order in $\beta$ and $J_{\tx{ex}}$,
\begin{align}
T = \frac{m^{\ast}}{\hbar}
\frac{\beta J_{\tx{ex}}}{E_{F}}\sg_{\tx{D}}
\end{align}
where $m^{\ast}$ is the effective mass. The Fermi energy
$E_{F}$ and the Drude conductivity are given by
\begin{align}
E_{F}=\frac{\hbar^{2}}{2m^{\ast}}(3\pi^{2}p)^{2/3},~
\sg_{D}=\frac{e^{2}\tau}{m^{\ast}}p,
\end{align}
where $\tau$ is the transport time. The last two relations immediately
give rise to $T\propto p^{1/3}$.
In the six-band model, the Fermi surface deviates from
a sphere and, as the value of $\beta$ increases,
the spin-orbit coupling starts to modify the density of states.
Both effects render the torque-versus-hole concentration curve
away from the $p^{1/3}$ dependence. This effect is illustrated in Fig. \ref{fig:st-hole}(b).
The former (strong spin-orbit coupling) clearly deviates from $p^{1/3}$,
whereas the latter (weak spin-orbit coupling) follows the expected $p^{1/3}$ trend.

In conclusion, in a DMS system subscribing to a linear Dresselhaus
spin-orbit coupling, we have found that the angular dependence of
the spin-orbit torque has a strong yet intriguing correlation with the
anisotropy of the Fermi surface. Our study also reveals a
nonlinear dependence of the spin torque on the exchange coupling.
From the perspective of material selection, for an equivalent set
of parameters, the critical switching current needed in (In,Mn)As
is expected to be lower than that in (Ga,Mn)As. The results
reported here shed light on the design and
applications of spintronic devices based on DMS.

Whereas the materials studied in this work have a Zinc-Blende
structure, DMS adopting a wurtzite structure, such as (Ga,Mn)N,
might also be interesting candidates for spin-orbit torque
observation due to their sizable bulk Rashba spin-orbit coupling.
However, these materials usually present a significant Jahn-Teller
distortion that is large enough to suppress the spin-orbit
coupling.\cite{GaN} Furthermore, the formalism developed here
applies to systems possessing delocalized holes and long range
Mn-Mn interactions and is not adapted to the localized holes
controlling the magnetism in (Ga,Mn)N.

We are indebted to K. V\'yborn\'y and T.
Jungwirth for numerous stimulating discussions.
F.D. acknowledges support from KAUST Academic Excellence
Alliance Grant N012509-00.

\end{document}